\pdfoutput=1
\documentclass[letterpaper]{article} 

\usepackage{graphicx} 
\usepackage{aaai23}  
\usepackage{times}  
\usepackage{helvet}  
\usepackage{courier}  
\usepackage[hyphens]{url}  
\usepackage{graphicx} 
\usepackage{natbib}  
\usepackage{caption} 
\usepackage{subfloat}
\usepackage{subcaption}
\usepackage{booktabs}
\usepackage{ragged2e}
\usepackage{blindtext}
\usepackage{algorithm}
\usepackage{algorithmic}
\usepackage{amsmath}
\usepackage{bm}
\DeclareCaptionLabelSeparator {fullstop} {.\quad
} 
\usepackage{newfloat}
\usepackage{listings}
\DeclareCaptionStyle{ruled}{labelfont=normalfont,labelsep=colon,strut=off} 
\floatstyle{ruled}
\newfloat{listing}{tb}{lst}{}
\floatname{listing}{Listing}

\usepackage{amsmath,amsthm,amscd,amssymb,amsbsy,mathtools}
\usepackage{amsfonts}
\usepackage{cancel}
\usepackage{enumitem}
\usepackage{array}
\usepackage{booktabs}
\usepackage{multirow}
\usepackage{bbding}
\usepackage{ulem}
\def\bbR{\mathbb R}

\title{ScatterFormer: Locally-Invariant Scattering Transformer for Patient-Independent Multispectral Detection of Epileptiform Discharges}

\author{
    Ruizhe Zheng\equalcontrib\textsuperscript{\rm 1,\rm 2,\rm 3,\rm 4,\rm 5,\rm 6},
    Jun Li\equalcontrib \textsuperscript{\rm 1,\rm 6}, 
    Yi Wang\thanks{Corresponding Author}\textsuperscript{\rm 7},
    Tian Luo\textsuperscript{\rm 7},
    Yuguo Yu\thanks{Corresponding Author}\textsuperscript{\rm 1,\rm 2,\rm 3,\rm 4,\rm 5,\rm 6}   
}

\affiliations{
    \textsuperscript{\rm 1} Research Institute of Intelligent and Complex Systems, Fudan University\\
    \textsuperscript{\rm 2} State Key Laboratory of Medical Neurobiology, Fudan University\\
    \textsuperscript{\rm 3} MOE Frontiers Center for Brain Science, Fudan University\\
    \textsuperscript{\rm 4} Institutes of Brain Science, Fudan University\\
    \textsuperscript{\rm 5} Institute of Science and Technology for Brain-Inspired Intelligence, Fudan University\\
    \textsuperscript{\rm 6} Shanghai Artificial Intelligence Laboratory\\
    \textsuperscript{\rm 7} Department of Neurology, Children’s Hospital of Fudan University\\
    rzzheng@fudan.edu.cn, jun\underline{\hbox to 0.1cm{}}li@fudan.edu.cn,  yiwang@shmu.edu.cn, luotian@fudan.edu.cn, yuyuguo@fudan.edu.cn
}

\usepackage{bibentry}

\begin{document}

\maketitle

\begin{abstract}
	Patient-independent detection of epileptic activities based on visual spectral representation of continuous EEG (cEEG) has been widely used for diagnosing epilepsy. However, precise detection remains a considerable challenge due to subtle variabilities across subjects, channels and time points. Thus, capturing fine-grained, discriminative features of EEG patterns, which is associated with high-frequency textural information, is yet to be resolved. In this work, we propose Scattering Transformer (ScatterFormer), an invariant scattering transform-based hierarchical Transformer that specifically pays attention to subtle features. In particular, the disentangled frequency-aware attention (FAA) enables the Transformer to capture clinically informative high-frequency components, offering a novel clinical explainability based on visual encoding of multichannel EEG signals. Evaluations on two distinct tasks of epileptiform detection demonstrate the effectiveness our method. Our proposed model achieves median AUCROC and accuracy of $98.14\%$, $96.39\%$ in patients with Rolandic epilepsy. On a neonatal seizure detection benchmark, it outperforms the state-of-the-art by $9\%$ in terms of average AUCROC.
\end{abstract}

\section{Introduction}

Continuous electroencephalography (cEEG) plays an important role in monitoring and diagnosis of epilepsy. Automatic detection of epileptiform discharges and distinguishing these activities from nonepileptiform abnormalities \cite{yum_shvarts_2019, shi2020differences} and normal/benign EEG activities \cite{li2020distinguishing} are of great importance in the biomedical field, because that experts with various levels of diagnostic experience usually report discrepant opinions on the EEG records \cite{xiang2015detection}. In particular, patient-independent epileptiform detection aims at identifying seizures within an EEG recording without separate fine-tuning efforts on new data to establish a subject-specific detector. However, pitfalls for correct identification of epileptiform discharges include misreading, misinterpretation and overinterpretation of individual EEG signature due to inherent subjectivity, can lead to misdiagnosis of people who do not have epilepsy \cite{tatum2012eeg, shorvon2016right, Tatum862}. Therefore, cost-effective, cross-subject algorithms are urgently in need.

So far, a plethora of deep learning-based algorithms based on visual spectral displays of EEG have been developed \cite{tatum2018clinical, rasheed2020machine, saminu2021recent}. Still, major concerns on the failure in clinical practice remain present due to limited generalizability of deep learning models on out-of-distribution data when conducting patient-independent predictions \cite{achilles2018convolutional, li2020distinguishing}, which requires precise identification of fine-grained features \cite{9385307} that is clinically interpretable for electroencephalographers with reasonable sensitivity and specificity. However, traditional learning schemes heavily depend on labor-intensive feature selection with regard to time-frequency subbands and electrodes, as well as often inconsistent preprocessing of raw data, which collectively limit the potential to discover a latent, informative representation of signals. Moreover, manual removal of intra- and inter- subject noises and artefacts could potentially eliminate features crucial for a fine-grained analysis of seizure occurrence.

Seizure dynamics is characterized by high-frequency outbursts of epileptiform abnormalities. Recent advances in attention-based deep neural networks, especially Transformers, have prompted augmented representational learning for EEG classification tasks \cite{bagchi2022eeg, siddhad2022efficacy,tao2021gated, sun2021eeg}. However, the problem of loss of relevant clinical biomarkers of irregularly altered EEG signals has not been explicitly addressed. Invariant scattering transform is a non-linear transform based on a cascade of wavelet transforms and modulus non-linearity. Invariant scattering convolution has energy preservation nature and eliminates translation and rotation variability algorithmically due to its Lipschitz-continuity \cite{mallat2012group}. Therefore, scattering coefficients are sensitive representation of complex, edge-like patterns \cite{cotter2019learnable, bruna2013invariant, mallat2012group}. These subtle patterns have been proven to be beneficial for texture classification \cite{singh2017dual}. More importantly, invariant scattering transform can retrieve high frequency components lost due to low-pass filtering by cascaded wavelet decomposition. Despite potential benefits, its application has not been investigated in Transformer learning of epileptiform recognition.

In this work, we aim at capturing subtle discriminative features of EEG data by strategically leveraging invariant scattering transform in Transformer architecture design. We apply a frequency-aware attention (FAA) to capture richer contextual dependencies, which incorporates scattering layers to prevent oversmoothing. The design allows the model to precisely capture fine-grained features using invariant scattering transform in combination with dynamic weighting of feature maps. An end-to-end multispectral diagnostic pipeline is established and rigorously tested in classifying spectra calculated directly from raw EEG records on one private and one public dataset. 
Furthermore, we present our correlation analysis of scattering transform, and show our theoretical result that scattering transform can improve the model generalizability. 
Our main contributions are summarized as follows:
\begin{itemize}
	\item We propose an end-to-end diagnostic pipeline targeting patient-independent epileptic seizure detection. The proposed approach is efficient in capturing fine-grained information in seizure-specific multispectral representations while maintaining time- and frequency-shift invariance, presenting a novel multispectral interpretability without compromising performance.
	\item We first introduce scattering transform to Transformer, which is accomplished through token embedding and frequency-disentangled attention based on locally-invariant scattering layers, leading to synergy between high- and low-frequency attention in order to obtain local discriminative patterns that is critical for epiletiform recognition without increasing computational load. 	
	\item We theoretically present our correlation analysis of wavelet transform, and show that the generalizability in patient-independent setting can be improved via reducing upper bound of Gaussian complexity.
	\item Extensive experimental comparisons based on cross-validation are evaluated over different epilepsy diagnosis datasets in a cross-subject manner. The results demonstrate that our model outperforms state-of-the-art methods.
\end{itemize}
\section{Related Work}
In this section, we briefly review related work on spectral detection of epileptic EEG activities, frequency-aware Transformers and invariant scattering transform, highlighting the recent advances and gaps.
\subsection{EEG Learning Representation}
\cite{asif2020seizurenet} proposed SeizureNet, a diagnostic framework based on multispectral fusion-based encoding of EEG data using short-time Fourier transform (STFT) of multichannel records and saliency mapping of STFT spectrograms, which is showed to be beneficial for capturing fine-grained information. The model generalizes well on unseen patient EEG epochs. In order to optimize DNN-based diagnostic method in terms of capturing fine-grained features as well as their long-range dependencies, \cite{9385307} proposed an attention-based model that separately captures global attention and fine-grained information for seizure detection. In particular, inductive bias is introduced through stacking of convolutional blocks, which are sensitive to high-frequency components, before self-attention calculation. \cite{bagchi2022eeg} introduced convolutional feature expansion to Transformer to model the inter-channel similarities. Different from previous research, we propose a mechanistically feasible design to help mitigate low-frequency bias that could cause performance degradation, and to obtain interpretability that highlights clinically informative attributes.
 
\subsection{High-Frequency Components and Attention Mechanism}
Transformers are capable of capturing low-frequency components, which are associated with global semantic information. Undesirable low-pass filtering occurs with depth increasing, which could potentially lead to over-smoothing of local textures, thus weakening the modeling capability. Recent works have introduced convolution operators to ViTs to strengthen capability of modeling local dependencies. Moreover, disentanglement of attention for parallel modelling of high- and low- frequencies not only alleviates over-smoothing, but also aggregates richer components that empirically bring performance gain. \cite{pan2022fast} proposed HiLo, which separately deals with low and high frequencies in a multi-head self-attention (MHSA) module to achieve disentanglement of feature maps in spectral domain at the same encoder layer without compromising computational efficiency.  \cite{si2022inception} proposed Inception Transformer, where high-frequency components are modelled by convolution and max-pooling operations to aggregate features across frequency range. In this work, ScatterFormer aims at more balanced modeling of high- and low- information in a more precise, interpretable manner by introducing frequency-aware operations that decompose and mix tokens in a way that preserves locally invariant high-frequency features.

\section{Preliminaries}
\subsection{Problem Formulation}
The EEG activities are recorded as multivariate time series $\\X = {\left\{{X}_i \right\}}^T_{i=1} \in\bbR^{T\times C}$, where $X_i = {\left\{x_{i,c}\right\}}_{c \in C} \in \bbR^{C}$ are recorded signals of multiple channels at the time point $i$, and $X$ represents an epoch of brain waveforms segmented manually or automatically with appropriate choice of duration $T$. Domain knowledge is employed to predetermine number of electrodes and EEG montage in order to obtain $C$ channels of signals. Preprocessed EEG epochs are encoded with the proposed saliency-aware multispectral representation. In real-word scenarios, two clinically relevant problems are defined for cross-subject detection of epileptiform discharges based on model $P_{\theta}$ parameterized with $\theta$.
\begin{itemize}
\item
Given $N$ clinically diagnosed subjects with and without epileptic seizures, $n_i$ EEG epochs and corresponding annotations are available for each individual, which constitutes the training dataset ${\left\{{\left\{X_k, y_k\right\}}^{n_i}_{k=1}\right\}}^{N}_{i=1}$, where $X_k \in\bbR^{T\times C}$, $y_k\in \{0, 1\}$ is the annotated label. For recently hospitalized $M$ patients who are yet to be diagnosed, $m_j$ EEG epochs are extracted for the $j^{th}$ individual, ${1\leq j\leq M}$, the estimate of the label is the probability of seizure occurrence:
\begin{equation}
\left\{\hat{y}^{seizure}_k\right\}^{m_j}_{l=1}=P_{\theta}(y|\left\{X_l\right\}^{m_j}_{l=1}, \left\{{\left\{X_k, y_k\right\}}^{n_i}_{k=1}\right\}^{N}_{i=1})
\end{equation}
\item 
Given $N$ clinically diagnosed subjects with epileptic seizures, $n_i$ EEG epochs and corresponding annotations are available for each individual, which constitutes the training dataset ${\left\{{\left\{X_k, y_k\right\}}^{n_i}_{k=1}\right\}}^{N}_{i=1}$. We aim to predict interictal and ictal activities during continuous EEG monitoring of out-of-domain subjects, who are recently diagnosed with epileptic seizures. For an EEG epoch $X \in\bbR^{T\times C}$, $y_k\in \{0, 1\}$ recorded at a specific moment, the estimate of the label is the probability of whether it is interictal or ictal:
\begin{equation}
\left\{\hat{y}^{ictal}_k\right\}^{m_j}_{l=1}=P_{\theta}(y|\left\{X_l\right\}^{m_j}_{l=1}, \left\{{\left\{X_k, y_k\right\}}^{n_i}_{k=1}\right\}^{N}_{i=1})
\end{equation} 
\end{itemize}

\subsection{Vision Transformer}
Vision Transformer (ViT) and its variants are highly capable of capturing global contextual information and long-range dependence in attention-based modeling of a plethora of modalities, including EEG spectra \cite{bagchi2022eeg, tao2021gated}, but are not capable enough to learn high-frequency components that are crucial for fine-grained information extraction and classification \cite{wang2022anti, bai2022improving}. Recent years have witness significant advance in examination of ViT from spectral domain, which contributes to resolving important gaps such as attention collapse \cite{wang2022anti}. Moreover, convolutional neural networks (CNNs) have been re-introduced into Transformer architecture to enhance sensitivity of multi-head self-attention (MHSA) to local features while maintaining reasonable computational load \cite{wu2021cvt}. In this work, we compute MHSA by cross-covariance attention proposed in \cite{ali2021xcit} as follows:
\begin{align}
    \boldsymbol{Attn}&=\bm{V} \mathrm{Softmax} \left(\frac{\bm{\hat{Q}}^{\top}\bm{\hat{K}}}{\tau} \right )
\end{align}
where $\boldsymbol{Attn}$ denotes attention maps, $\bm{\hat{Q}}$, $\bm{\hat{K}}$ are $l_2$-normalized query and key embeddings, $\bm{V}$ is value embedding, $\tau$ is a learnable parameter for stabilizing training.
\section{Methodology}
\subsection{Saliency-Aware Multispectral Representation}
We propose a novel method to generate a saliency-aware, multi-spectral, multi-channel representation of the EEG records shown in bipolar montage, where the localization of the celebral potential is based on the direction of the waveform between two channels, and a phase reversal is beneficial for easier identification of epileptiform abnormalities \cite{sazgar2019overview}. Specifically, continuous wavelet transformation (CWT) to circumvent the problem of non-stationarity.
\begin{equation}
	W_\psi[x(a,b)] = \int_{-\infty}^{+\infty}x(t)\bar{\psi}(\frac{t-b}{a})dt\label{eq}
\end{equation}
where $\psi(\cdot)$ denotes a family of base functions that dilate and contrast with frequency to analyze intricate time-frequency details.
Power spectrum $S_i=log(|W_{\psi_i}[x(a,b)]|^2)$ in terms of mother wavelet $\psi_i(\cdot)$. The multispectral representation is formulated as $S=norm(\sum_{i}S_i)$, where $norm(\cdot)$ denotes $l_2$ normalization. The spectrogram is stacked with static spectral saliency map $SA_{1}$ and fine-grained saliency map $SA_{2}$ to incorporate more abundant information that are integral to diagnosis. Differences of Gaussian (DoG), Paul and Morlet are used to compute $S$. 
\subsection{Invariant Scattering Transformer}
We propose frequency-aware attention that employs dual branches of attention calculation and local sensitivity of invariant scattering transform to disentangle high- and low-frequencies. 

\subsubsection{Invariant Scattering Token Embedding}
In order to accomplish efficient identification and encoding of information that is typically lost in the down-sampling of high-resolution fine-grained multispectral visual representation, invariant scattering operator, which is Lipschitz-continuous to diffeomorphisms that causes significant perturbations to high-frequency components, is proposed to preserve information without over-smoothing.
An invariant scattering transform provides locally translation-invariant multiscale coefficients, which characterize the scaling properties of signals. They are computed by iteratively calculating the modulus of complex wavelet coefficients, which are yielded by convolving input signal with mother wavelet $\psi(\cdot)$ and scaling function $\phi(\cdot)$. A dyadic band-pass filter bank is determined for $j \in\mathbb{Z}$ and rotation $r \in G$, $G$ is a finite rotation group in $\bbR^2$.
\begin{equation}
\psi_{2^jr}(x)=2^{2j}\psi(2^jr^{-1}x)
\end{equation}

The 2-D wavelet transform is done by convolving the input with a mother wavelet
dilated by $2^j$ and rotated by $\theta$:

\begin{equation}
  \psi_{j, \theta}(x) = 2^{-j}\psi \left(2^{-j} R_{-\theta} x\right)
\end{equation}

where $R$ is the rotation matrix, $1 \leq j \leq J$ are indexes of the scale.
We define $ \Lambda_{J}^1 := \{( j, r) ~:~ 1 \leq j \leq J ~\text{and}~ r \in G\}$. For $p := (\lambda_{1} , \lambda_{2}, \cdots, \lambda_m) \in \Lambda_{J}^{m} := \Lambda_{J}^1 \times\Lambda_{J}^1\times\cdots \times \Lambda_{J}^1$ ($m$ times), a scattering propagator $U[p] : L^{2}(\bbR^2) \to L^{2}(\mathbb{R}^2)$ is defined as

\begin{equation}
	U[p]f(x):=U[\lambda_{m}]\cdots U[\lambda_{2}]U[\lambda_1]f(x) ~~ \forall f \in L^{2}(\mathbb{R}^2),
\end{equation}
where $U[\lambda_k]f(x) := |(\psi_{\lambda_k} * f)(x)|$ for $k = 1, 2, \cdots, m$ and $U[\emptyset]f(x) := f(x)$.
We compute $m$ times of convolutions and modulus operators along the path of $p$ of length $m$, the input $f$ is propagated to the $m$-th layer:
\begin{equation} 
S_J[p]f(x) := (\phi_{2^J} * U[p]f)(x).
\end{equation}
The resulting $ m^{th} $-order scattering coefficients has energy preservation property. Specifically, We define scattering decomposition of input $f(x)$ as
\begin{equation} 
S_J[\Lambda_{J}^1] f(x) := \{S_J[p]f(x)\}_{p \in \Lambda_{J}^1} 
\end{equation}
Its norm is $\sum_{p\in \Lambda_{J}^1} \|S_J[p] f(x)\|^2$. Note that 
$ S_J[\Lambda_{J}^1] $ is contractive:
\begin{equation}
\|S_J f(x) - S_J f(y) \| \leq \|f(x) - f(y)\|~.\
\end{equation}
It is further proved that \cite{mallat2012group}

\begin{align}
\lim_{m_{\max} \rightarrow \infty} \sum_{m=m_{\max}}^{\infty} \|S_J [\Lambda_J^m] x \|^2 &= 0~.
\end{align}

The above result implies that invariant scattering transform is energy-preserving in addition to its stability to diffeomorphisms \cite{mallat2012group, singh2017dual}. Recent approaches have indicated that hybrid neural networks with scattering layers can achieve competitive performance on a range of benchmarks such as ImageNet and CIFAR-10 \cite{cotter2019learnable, singh2018generative, singh2017dual} at a faster convergence rate. Unlike CNNs, it has not been strategically introduced to Transformer to improve the effectiveness of the Transformers. 
Typically, second-order invariant scattering transform with optional learnable channel-wise mixing of maps modeled by conventional convolution operation \cite{cotter2019learnable}. However, \cite{oyallon2018compressing} noted that first-order scattering alone can effectively preserve discriminative information. In light of this, we adopt a hierarchical design in ScatterFormer by applying multiple invariant scattering layers to incorporate more high-frequency structures.
\begin{figure*}[!t]
\centering
\includegraphics[width=1.0\linewidth]{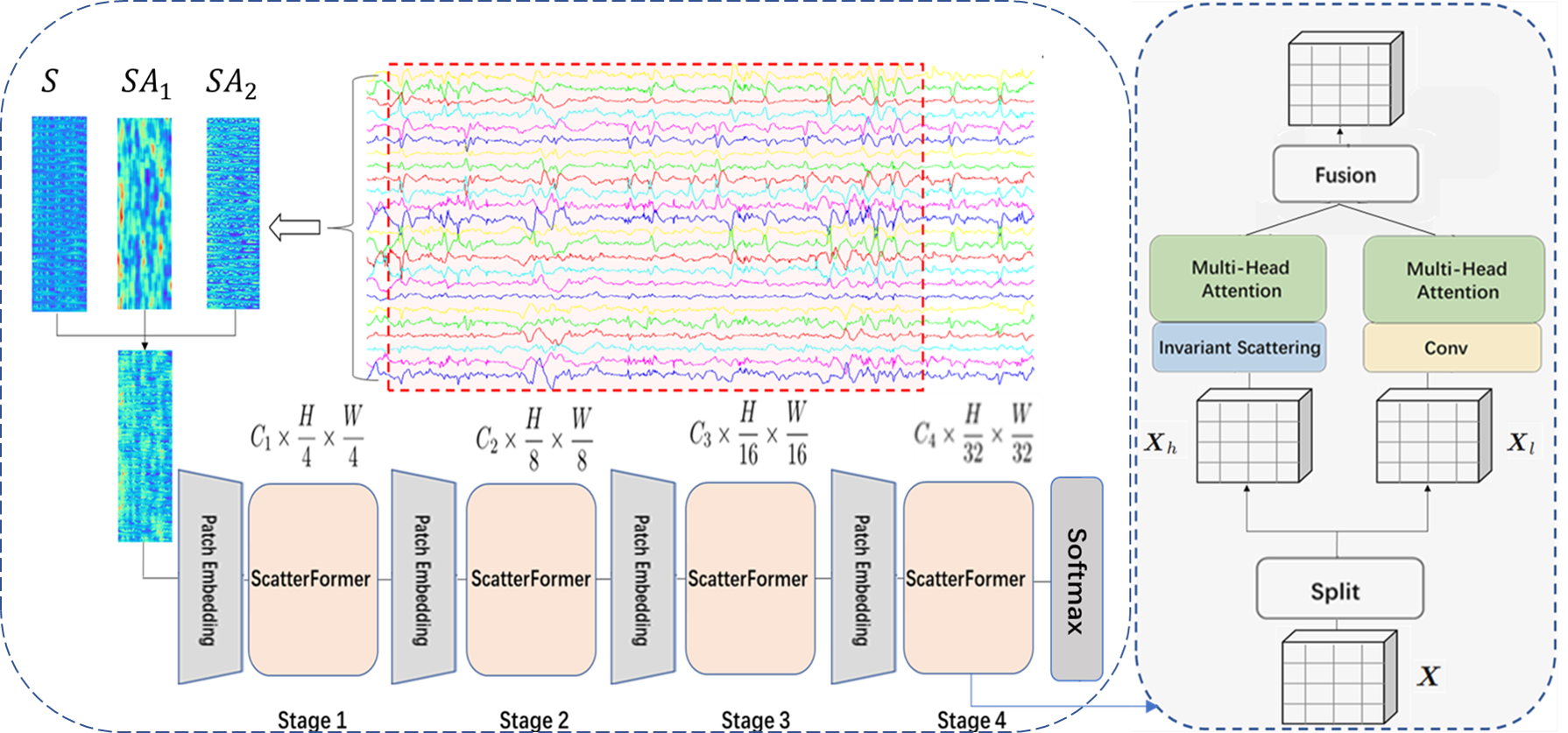}
\caption{Left: Architecture of ScatterFormer and diagnostic pipeline based on multispectral visual representation of EEG epochs for epileptiform identification. Input features are the raw EEG signals processed by multispectral representation. On top of the hierarchical stacking of ScatterTransformers is a Softmax classifier to predict epileptiform discharges.Right: Design of frequency-aware attention (FAA). The convolution (Conv) encoding of low-frequency tokens and invariant scattering encoding of high-frequency tokens are processed in separate pipelines. Attention maps are followed by channel mixing using pointwise convolution. The outputs are fused feature representations. Within the multi-head attention, we utilize cross-covariance attention to reduce computational cost.}
\label{fig:framework}
%
\end{figure*}

\subsubsection{Frequency-aware Attention (FAA)}
The tokens are split apart before being separately processed in dual-branch attention module. For high-frequency branch, we propose to use an invariant scattering transform projects feature maps to queries $\boldsymbol{Q}$. Multiple invariant scattering layers located at the beginning of each stage acting as down-sampling and patch-merging module together with those reside within the encoder intuitively preserves high-frequency scattering coefficients required to discriminate between images. Moreover, first-order transform is able to capture important attributes with minimal energy loss \cite{bruna2013invariant,oyallon2018compressing}. Thus, higher-order coefficients could offer only marginal gains while significantly increasing memory and computational cost in the high-frequency branch of self-attention.
In conformity with reduced resolution of token maps, the projection of keys $\boldsymbol{K}$ and values $\boldsymbol{V}$ uses $3\times 3$ convolution with a stride of $2\times 2$, which further reduces the computational cost. The high-frequency attention is formally expressed as
\begin{align}
    \boldsymbol{Q}_h&= \mathrm{BatchNorm}\left(\mathrm{Inv} \left( \boldsymbol{X}_h  \right) \right) \\
     \boldsymbol{K}_h&= \mathrm{BatchNorm}\left(\mathrm{Conv}_K \left( \boldsymbol{X}_h  \right) \right) \\
    \boldsymbol{V}_h&= \mathrm{BatchNorm}\left(\mathrm{Conv}_V \left( \boldsymbol{X}_h  \right) \right)
\end{align}
for $\boldsymbol{X}_h \in \bbR^{\frac{C_i}{2}\times H_i\times W_i}$ at $i^th$ stage, where $\mathrm{Inv}$ denotes invariant scattering layer, $\mathrm{Conv}_K$ and $\mathrm{Conv}_V$ denote convolution, $\mathrm{BatchNorm}$ denotes batch normalization, which is used to stablize the training procedure. Following \cite{ali2021xcit}, $\boldsymbol{Q}_k$ and $\boldsymbol{K}_h$ are normalized to obtain channel-wise attention score matrix. 
\begin{align}
    \bm{X}_h&= \bm{V}_h\mathrm{Softmax}\left( \bm{Q}^{\top}_h \bm{K}_h\right) 
\end{align}
Low-frequency attention is calculated in the separate path without  Transformer encoder is demonstrated to be inherently sensitive to low-frequency components \cite{wang2022anti}. We use $3\times 3$ convolution for token embedding. Linear projection layer for two groups of attention maps is used to mixing the separately learned attention in order to obtain a common representation. The locally-enhanced relative position encoding (LePE) \cite{dong2021cswin} is applied on $\boldsymbol{X}\in \mathbb{R}^{\frac{C_i}{2}\times H_i\times W_i}$ before attention calculation and added on the fused attention maps $\boldsymbol{X}_{fused}$:

\begin{align}
    \boldsymbol{X}_{fused}&= \mathrm{Linear}(\mathrm{Concat}(\boldsymbol({X}_l), \mathrm{Upsample}(\boldsymbol{X}_h)))\\
    \boldsymbol{X}_{fused}&=\boldsymbol{X}_{fused}+\mathrm{DWConv}(\boldsymbol{X})
\end{align}
where $\mathrm{Linear}$ denotes linear projection, $\mathrm{Upsample}$ denotes bilinear upsampling module, $\mathrm{DWConv}$ denotes depthwise convolution, $\mathrm{Concat}$ denotes concatenation.

\subsection{Generalization Capacity Analysis}
In this section, we theoretically prove that invariant scattering transform can enhance the generalization capacity of a network.

We first introduce the Gaussian complexity, 

\begin{align}\label{eq:Gaussian complexity}
\hat{G}_N(F) = \mathbb{E}[sup_{f \in F} \frac{2}{N}\sum_{i=1}^{N}g_if(x_i)]
\end{align}
where $g_i, i=1,2,\cdots,N$ are independent standard normal random variables. Gaussian complexity in Eq.~\eqref{eq:Gaussian complexity} measures the capacity of a functional class $F$. Gaussian complexity (and the closely-related Rademacher complexity) is often linked with the generalization analysis of deep learning, as its definition does not rely on the number of model parameters.

Then, we link correlation analysis that examines locality of features with the generalization capacity of transformer. As recent work \cite{NEURIPS2021_c404a5ad} shows that increase locality of features can compensate the high-frequency features in transformers, strong feature correlation is expected to improve the generalization capacity of transformers. The following theorem shows that feature correlation is closely related to an upper bound of Gaussian complexity.


\textbf{Theorem 1 [\cite{li2017filter}]} Suppose that $\sigma:\mathbb{R}\to \mathbb{R}$ is a contraction mapping. Define the class computed by one convolutional layer followed by one fully connected layer with 2-norm constraint as:
\begin{align}
    F = \{\bm{x} \to \sum_{i} v_i \sigma (\bm{w}_i)\bm{x}:||\bm{v}||^2 \le 1, ||\bm{w}||_1 \le B\}
\end{align}
For any $\bm{x}_1,\bm{x}_2,\cdots,\bm{x}_N \in \mathbb{R}^d$, we have
\begin{align}
    \hat{G}_N(F) \le \frac{cB(\ln d)^{1/2}}{N} \max_{\bm{j}-\bm{j}'\in \mathcal{N}}\sqrt{\sum_{i=1}^{N}||\bm{x}_i(\bm{j})-\bm{x}_i(\bm{j}')||^2}
\end{align}
where $\mathcal{N} \subset \mathbb{Z}^p$ defines the shape of the convolution filter as:
\begin{align}
    (\bm{w}_i * \bm{x})(\bm{k})=\sum_{\bm{j} \in \mathcal{N} } \bm{w}_{i,\bm{j}}\bm{x}[\bm{k}](\bm{j}) 
\end{align}
where $\bm{j}$ is the integer index vector that denotes the index shift.
  It is easy to see that minimize 
$\sum_{i=1}^{N}||\bm{x}_i(\bm{j})-\bm{x}_i(\bm{j}')||^2$ is equivalent to maximize the feature correlation $cov(\bm{x}_i(\bm{j}),\bm{x}_i(\bm{j}'))$. Therefore, strong feature correlation can improve the network generalization in some sense. In the following, we show that Scatter transform or Fourier transform can increase feature correlation, thus may improve the model generalizability.






\subsubsection{Generalization Capacity Analysis of Wavelet Transform}
When tackling with scattering transform, we characterize it in the case of Morlet wavelet $\psi$:
\begin{align}\label{eq:Morlet}
    \psi(\bm{x}) = \|C_1(e^{i\bm{x}\xi}-C_2)e^{-|\bm{x}|^2/(2\sigma^2)}\|
\end{align}
The following theorem shows that Morlet wavelet $\psi$ can also increase the feature correlation.

\textbf{Theorem 2} ~ Suppose the input feature $\bm{x}$ is normalized, i.e. $||\bm{x}||=1$, then Morlet wavelet in Eq.~\eqref{eq:Morlet} can increase feature correlation by appropriately choosing $C_1, C_2, \xi$.

\begin{proof}\renewcommand{\qedsymbol}{}
\begin{align}
     &\|\psi(\bm{x}_i(\bm{j}))-\psi(\bm{x}_i(\bm{j}'))\| \nonumber \\
    \le& \|C_1\|\cdot\|(e^{i\bm{x}_i(\bm{j})\xi}-C_2)e^{-|\bm{x}_i(\bm{j})|^2/(2\sigma^2)} \nonumber\\
    -& (e^{i\bm{x}_i(\bm{j}')\xi}-C_2)e^{-|\bm{x}_i(\bm{j}')|^2/(2\sigma^2)}\| \nonumber\\
    \le&\|C_1\|\big(\|i\xi e^{i\bm{\eta}\xi} e^{-|\bm{\eta}|^2/(2\sigma^2)}\|\nonumber
    \\&+ \|(e^{i\bm{\eta}\xi} -C_2)e^{-|\bm{\eta}|^2/(2\sigma^2)}\frac{|\bm{\eta}|}{\sigma^2}\|\big) \nonumber \cdot |\bm{x}_i(\bm{j})-\bm{x}_i(\bm{j}')| \nonumber \\
    \le & \|C_1\|\big(\|\xi\| + (\|C_2\|+1)\frac{2}{\sigma^2} \big) \|\bm{x}_i(\bm{j}) - \bm{x}_i(\bm{j}')\|
\end{align}
As long as $\|C_1\|\big(\|\xi\| + (\|C_2\|+1)\frac{2}{\sigma^2} \big) < 1$, we can increase the feature correlation, thus reduce the upper bound of Gaussian complexity, as shown in Theorem 1.

\end{proof}

\subsubsection{Generalization Capacity Analysis of Fourier Transform}
We further analyze the generalization capacity of Fourier transform. Suppose a Fourier transform $\gamma$ featurize input coordinate with a set of sinusoids as follows:
\begin{align}\label{eq:Fourier}
    \gamma(\bm{x}) =& [a_1cos(2\pi \bm{b}_1^{\top}\bm{x}),a_1sin(2\pi \bm{b}_1^{\top}\bm{x}), \cdots, \nonumber\\ &a_mcos(2\pi \bm{b}_m^{\top}\bm{x}),a_msin(2\pi \bm{b}_m^{\top}\bm{x})]
\end{align}
where $m$ is dimension of input feature $x$, $a_i,b_i$ are parameters in the Fourier transform $\gamma$. 

Then the inner product of Fourier features can be represented as follows:
\begin{align}
    \gamma(\bm{x}_1)^{\top}\gamma(\bm{x}_2) = \sum_{k=1}^{m} a_k^2 cos \big(2\pi \bm{b}_k^{\top}(\bm{x}_1-\bm{x}_2)\big)
\end{align}

The following theorem shows that Fourier features can increase the feature correlation. 

\textbf{Theorem 3} ~ Suppose the input feature $\bm{x}$ is normalized, i.e. $||\bm{x}||=1$, then the Fourier transform in Eq.~\eqref{eq:Fourier} can increase feature correlation by appropriately choosing $a_1,a_2,\cdots,a_m$. 

\begin{proof}\renewcommand{\qedsymbol}{}
\begin{align}
    &\sum_{i=1}^{N}||\gamma(\bm{x}_i(\bm{j}))-\gamma(\bm{x}_i(\bm{j}'))||^2 \nonumber\\
    =& \sum_{i=1}^{N} \Big[\sum_{k=1}^{m}2a_k^2 - 2a_k^2 cos\big(2\pi \bm{b}_k^{\top}(\bm{x}_i(\bm{j})-\bm{x}_i(\bm{j}'))\big)\Big] \nonumber\\
   \le& \sum_{i=1}^{N}\Big[ \sum_{k=1}^{m}4a_k^2 \Big]
    = 4N\sum_{k=1}^{m}a_k^2
\end{align}

Therefore, we can increase the feature correlation by selecting $a_1,a_2,\cdots,a_m$ such that
\begin{align}
    4N\sum_{k=1}^{m}a_k^2 < \sum_{i=1}^{N}||\bm{x}_i(\bm{j})-\bm{x}_i(\bm{j}')||^2
\end{align}

\end{proof}
According Theorem 1, Fourier transform can reduce the upper bound of Gaussian complexity, thus may improve the model generalization capacity.


\section{Experiments}
\subsection{Model Architecture}
ScatterFormer adopts a hierarchical architecture illustrated in Figure \,\ref{fig:framework}, which is demonstrated to be effective for a range of tasks on two-dimensional signals due to its capability to generate multi-scale representation. Several variants are investigated. Specifically, main variants denoted by ConvScatter-1, ConvScatter-2 and ScatterFormer are examined, which correspond to networks with convolutional token embedding, with scattering token embedding positioned at the initial stage, with scattering token embedding positioned at all stages, respectively. In particular, the initial token embedding achieves $4\times4$ downsampling using a second-order invariant scattering layer. At later stages, a first-order invariant scattering layer is used. In order to investigate the frequency characteristics of ScatterFormer, we establish FourierFormer, where scattering layers are replaced with Local Fourier Unit (LFU). Moreover, we establish ProtoFormer, which calculates MHSA using naive setting of cross-covariance attention.
\subsection{Experimental Setup}
\subsubsection{Datasets}
Extensive experiments are performed on two datasets for evaluation of proposed diagnosing models, with performance on epileptiform discharges detection and neonatal seizure or ictal epileptiform discharges detection being evaluated, respectively. 

BECTS/Rolandic epilepsy dataset is a private dataset comprised of 110 patients (average age: $133.7\pm 27.4$ months) with BECTS/Rolandic epilepsy, a common child epilepsy, and 170 normal controls (average age: $131.6\pm25.3$ months). Collection and use of the data was approved by the Ethics Review Committee of the Children Hospital of Fudan University (No. 522-2020), and all subjects provided written informed consent. The annotated EEG epochs eligible for this work last for 446.55 h. Preprocessed data will be available at https://github.com/albertcheng19/scatterformer.

Helsinki University dataset is a public dataset that contains 39 patients with consensus annotation, 22 patients that are diagnosed as free of seizure in their records and 18 patients with no consensus annotation of ictal epileptiform discharges. Additional description can be referenced in \cite{stevenson2019dataset}. Data is available at https://zenodo.org/record/4940267.
\subsubsection{Data Collection and Preprocessing}
Routine EEG recordings were performed on all subjects underwent continuous ($>3$ hours) sleep state with Nicolet EEG system (Natus Medical, Incorporated, San Carlos, CA, USA) with Ag/AgCl electrodes in the EEG examination room at the Hospital. Patients were seizure free $\leq1$ hours before study and receiving stable doses of medication. EEG electrodes were placed according to the conventional 10–20 EEG system and the guideline of American Clinical Neurophysiology Society. The impedance of the electrodes was calibrated under 3 $k\Omega$. The EEG signals were amplified and digitized at a sampling rate of 250 Hz, and then filtered at 0.1 Hz high-pass, 100 Hz low-pass and notch filter of 50 Hz. The classical Rolandic epileptic and normal EEG periods were labeled by two senior clinical neurophysiologists. The raw EEG data were saved in EDF format. EEG signals were preprocessed to remove the linear trend and eye movement artifacts using independent component analysis (ICA). Bipolar montage is adopted in our experiments. 
\subsubsection{Regularization}
Two types of augmentation is applied in training to alleviate over-fitting. The first one is termed channel reshuffling in our work. The feature representation of each channel is resized into a $3\times32\times256$ array, and 24 multi-spectral feature arrays in total are randomly rearranged into a $3\times768\times256$ array with different channels located in different positions of the array. The proposed method excludes the potential bias introduced by the specific arrangement of channels. After reshuffling, the input data are subject to further augmentation using MixUp \cite{zhang2017mixup}. 
\subsubsection{Training Details} AdamW optimizer with weight decay of 0.05 is used. The initial learning rate is set to 5e-4 and progressively decays after each interaction by a cosine scheduler. During training, exponential moving average (EMA) with decay at 0.9999 is used to smooth the updating of weights. Unless otherwise stated, all models are trained with an $768\times256$ input size. The maximum training epochs are set at 50 with early stopping. All experiments are conducted on 2 NVIDIA A100 GPUs.

\subsection{Experimental Results}
Evaluation results on BECTS/Rolandic dataset are reported in Table\;\ref{tb:results_1} (a). Results of several variants of ScatterFormer is presented. Trained with the same initialization, data augmentation, learning rate tuning and regularization, the results suggest that both patch embedding implemented with invariant scattering layer and convolutional layer could improve the performance, but applying scattering transform only at the first stage leads to drop in AUCROC. In particular, ScatterFormer outperforms other variants and a strong baseline CNN model. Results on Helsinki University dataset are reported in Table\;\ref{tb:results_1} (b). In comparison to previous work, our method significantly surpasses the state-of-the-art given the constraints on data availability and cross-validation. Additional evaluation is reported in Table 3. ScatterFormer achieves satisfactory sensitivity to epileptic EEG segments in patient-independent classification. Moreover, as can be observed in Figure 2, performance on different folds follows skewed distribution. More concentrated distribution of ScatterFormer suggests a more robust overall performance across all folds. Therefore, ScatterFormer is a more robust model for generalization on heterogeneous individuals.
\begin{table}[t]
\footnotesize
\centering
\resizebox{0.9\columnwidth}{!}{
\begin{tabular}{p{5em}|p{2.9em}lp{2.9em}lp{2.3em}lp{2.4em}}
\toprule[1pt]
\multicolumn{5}{c}{\textbf{(a) Results on BECTS/Rolandic Dataset}} \tabularnewline
  Model & Params (M) & LT ($\mu s$)& AUCROC & ACC \tabularnewline
\hline
\hline
     RegNet-Y-8G \\
     \cite{radosavovic2020designing} &37 & 179 & $94.52_{2.22}$& $88.11_{18.98}$\tabularnewline
     Swin-B
     \cite{liu2021swin}
     & 88 & 540 &$97.81_{0.65}$ & $94.04_{1.21}$ \tabularnewline
\hline
     ProtoFormer &   24 &79& $97.91_{3.45}$ & $92.87_{21.82}$\tabularnewline
     ConvScat-1 &   39& 300&  $98.15_{1.59}$&$96.33_{3.55}$  \tabularnewline
     ConvScat-2 &   39&  441& $96.85_{0.84}$&$95.10_{1.70}$  \tabularnewline
     ScatterFormer &   42&  365 &$98.14_{2.04}$&$96.87_{2.46}$  \tabularnewline
\bottomrule[1pt]
\end{tabular}}
\resizebox{0.9\columnwidth}{!}{
\begin{tabular}{p{5em}|p{2.9em}lp{2.9em}lp{2.3em}lp{2.4em}}
\toprule[1pt]
\multicolumn{5}{c}{\textbf{(b) Results on Helsinki University Dataset}} \tabularnewline
  Model & Params (M) & LT ($\mu s$)& AUCROC & ACC\tabularnewline
\hline
     ScatterFormer &  42&  365&  $96.38_{5.66}$&$90.55_{11.75}$ \tabularnewline
     SVM\\
     \cite{isaev2020attention} & - & - & $92.3_{12.1}$ & -  \tabularnewline
     \hline
     ScatterFormer & 42 & 365 & $90.31$ & $89.67$ \tabularnewline
     SWT-FCN\\
     \cite{9140713} & - & - &81 & 82 \tabularnewline
     SWT-CNN\\
     \cite{9140713} & - & -& 77 & 79 \tabularnewline
\bottomrule[1pt]
\end{tabular}
}
\caption{Results of ScatterFormer on patients with BECTS/Rolandic epilesy (a) and neonatal seizures (b). Compared with the existing deep learning models with similar amount of parameters, our method achieved superior outcomes and moderate inference time. We report evaluation metrics using median with interquartile range (IQR) as subscript. In addition, to compare our results of previous work, we also report values using mean value on Helsinki University dataset. Latency (LT) and number of parameters (Params) are reported for our experiments. }
\label{tb:results_1}
\end{table}
\begin{table}[htbp]
	\footnotesize
	\centering
	\resizebox{0.55\columnwidth}{!}{
		\centering
		\begin{tabular}{l|c c}
			\toprule[1pt]
			\multicolumn{3}{c}{\textbf{(a) BECTS/Rolandic Dataset}}
			\tabularnewline
			Model & AUCPR & F1\tabularnewline
			\hline
			\hline
			Proto & $98.18_{0.84}$ &$89.12_{0.82}$\tabularnewline
			Scatter &$98.88_{1.13}$&$93.93_{6.42}$\tabularnewline
			Fourier & $97.44_{4.72}$ & $86.26_{9.75}$\tabularnewline 
			\bottomrule[1pt]
	\end{tabular}}
	\resizebox{0.55\columnwidth}{!}{
		\begin{tabular}{l|c c}
			\toprule[1pt]
			\multicolumn{3}{c}{\textbf{(b) Helsinki University Dataset}}
			\tabularnewline
			Model & AUCPR & F1\tabularnewline
			\hline
			\hline
			Scatter & $92.32_{22.63}$ & $79.90_{28.88}$\tabularnewline
			Fourier &  $89.96_{37.35}$ & $70.05_{35.21}$\tabularnewline 
			\bottomrule[1pt]
	\end{tabular}}
	\label{tb:results_2}
	\caption{Results of AUCPR and F1-score metrics on two datasets. ScatterFormer achieves significantly higher performance, indicating improved sensitivity to ictal samples.}
\end{table}
\begin{figure}[!t]
	\centering
	\subfloat[BECTS/Rolandic Dataset]{
		\includegraphics[width=0.94\linewidth]{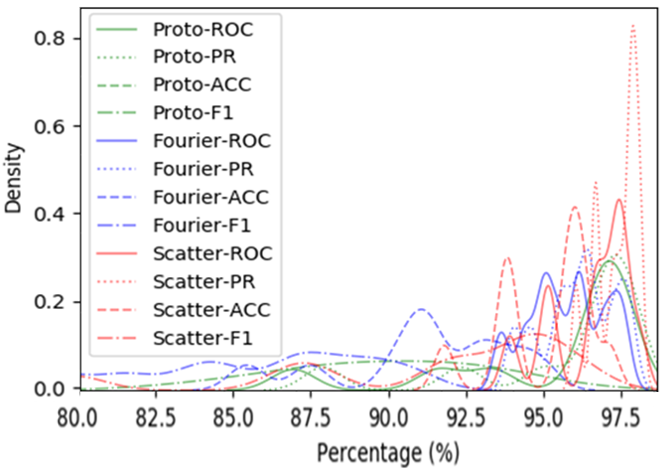}}
	\quad
	\subfloat[Helsinki University Dataset]{
		\includegraphics[width=0.94\linewidth]{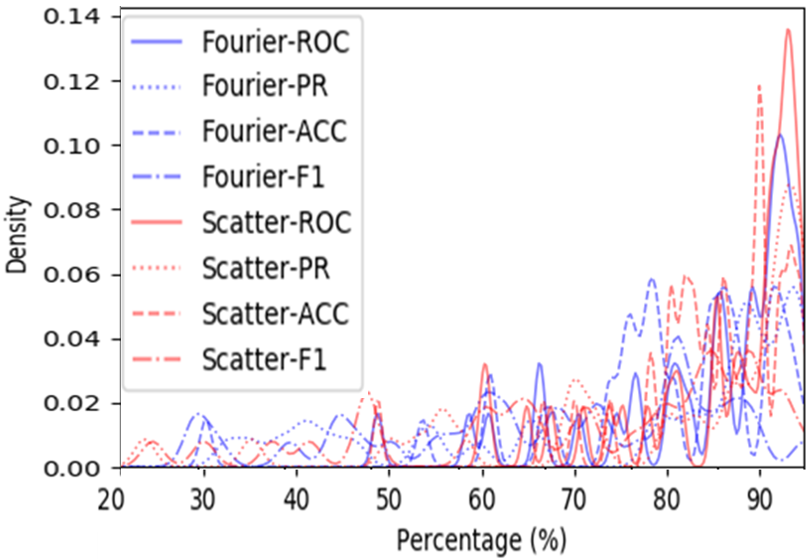}}
	\label{fig:distribution}
	\caption{Probability distribution of evaluation metrics across cross-validation folds. Density is estimated using kernel method. ScatterFormer have more concentrated probability distribution of various metrics on cross-validation folds. The results suggest that ScatterFormer achieves higher performance and generalizability than other models in cross-validation setting.}
\end{figure}
\begin{figure}[!t]
	\centering
	\subfloat[ScatterFormer]{
		\includegraphics[width=0.95\linewidth]{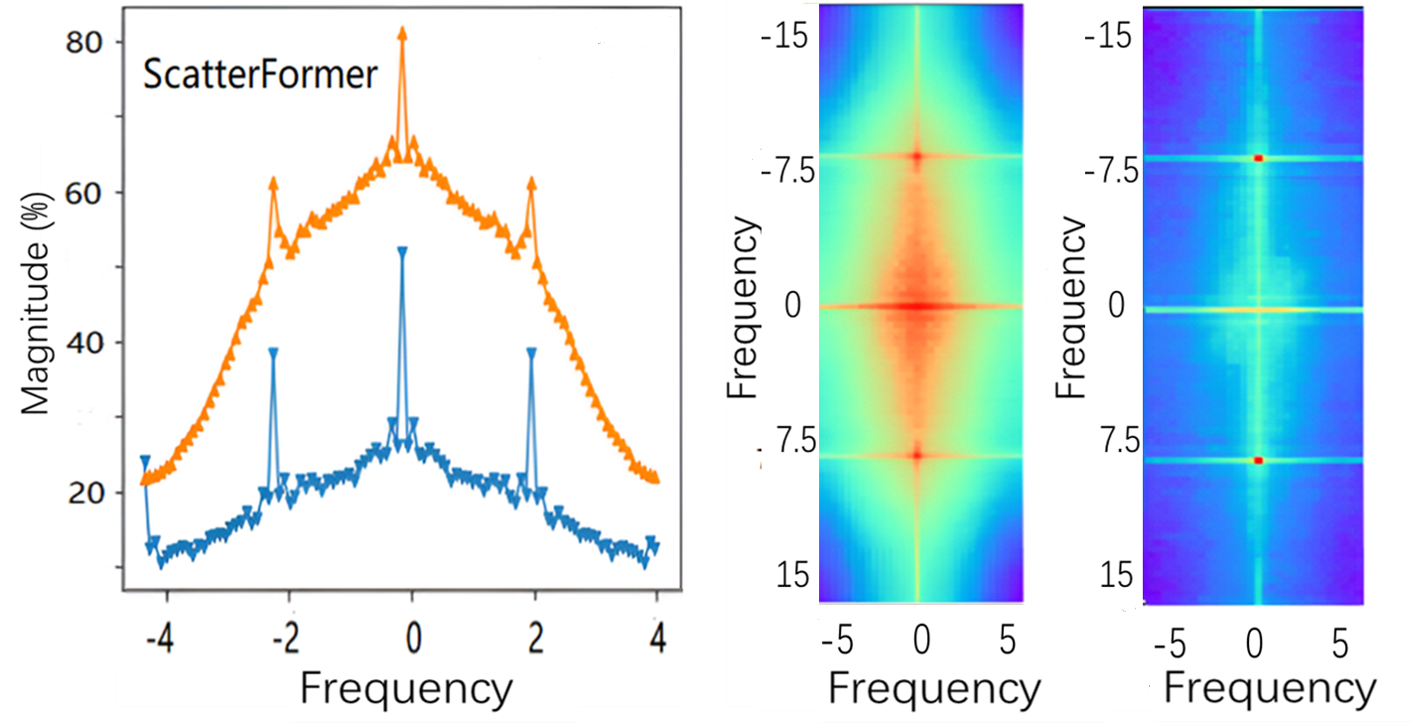}}
	\quad
	\subfloat[FourierFormer]{
		\includegraphics[width=1.0\linewidth]{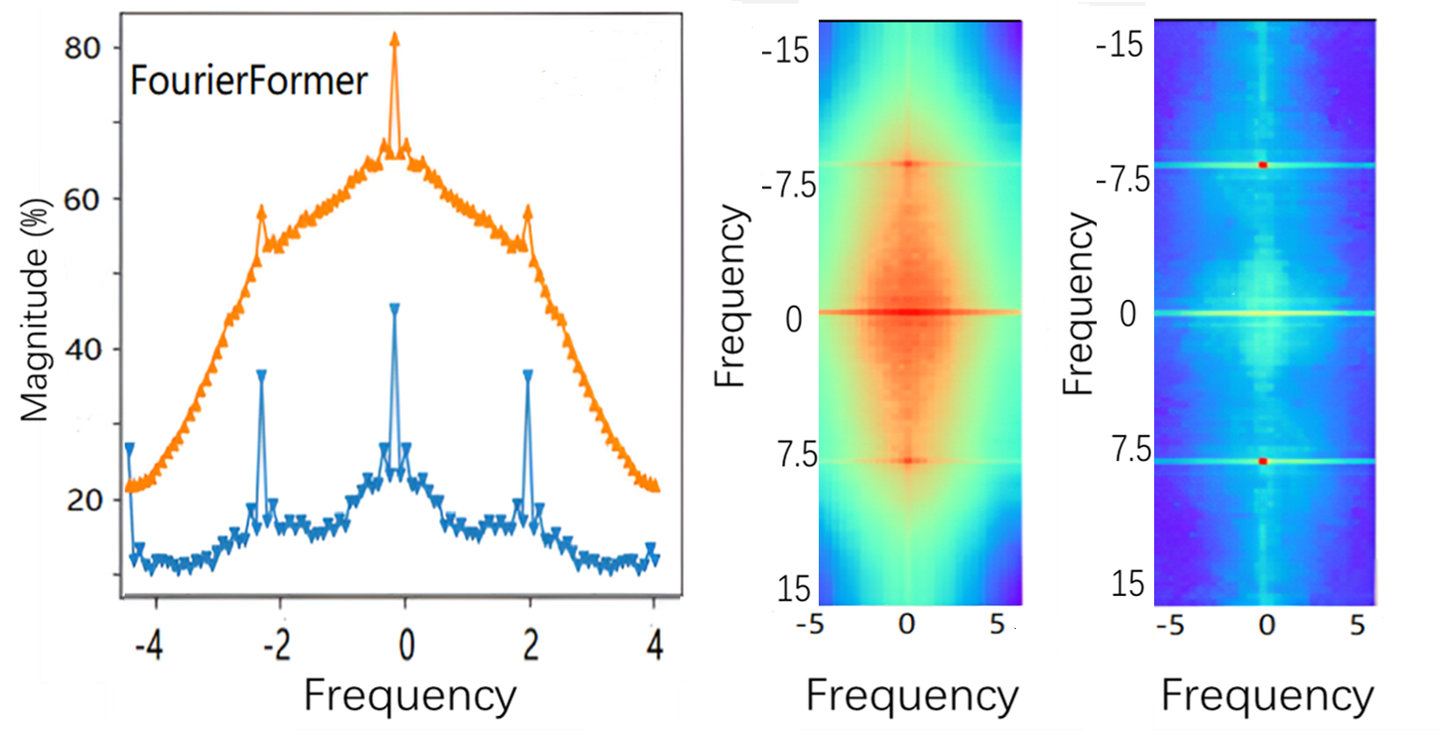}}
	\quad
	\subfloat[ProtoFormer]{
		\includegraphics[width=1.0\linewidth]{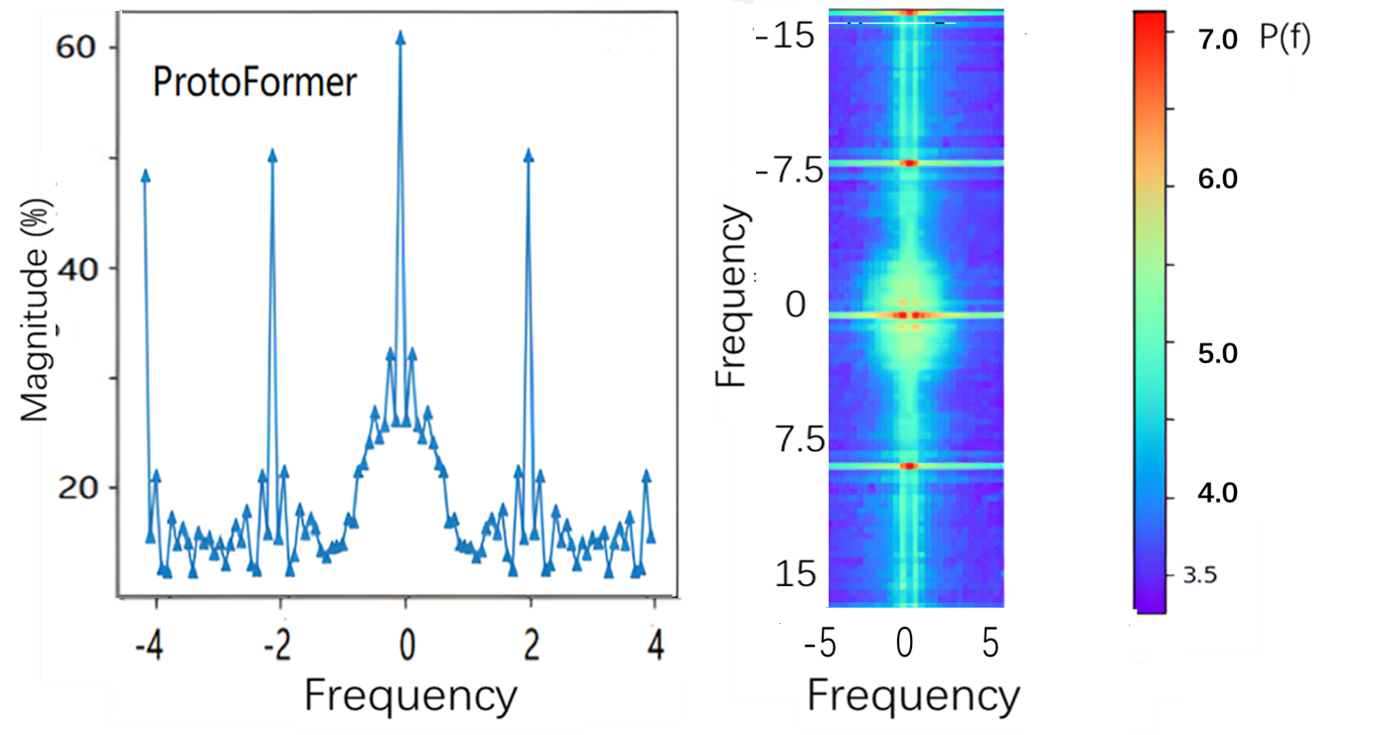}}
	\label{fig:fourierspectrum}
	\caption{Fourier spectrum analysis of attention maps of three main variants investigated in this work. The results are averaged across 100 random samples at the $4^{th}$ layer from a randomly selected attention head. Both invariant scattering (a) and fast Fourier convolution (b) reduces the low-frequency preferability observed in (c), where prototypical attention tends to have sharper frequency response. Magnitude-Frequency responses (d) show that high-frequency components are enhanced by the dual-branch design. Spectra of high- and low-frequency branches are denoted by yellow and blue curves, respectively, in ScatterFormer and FourierFormer.}
\end{figure}
\begin{table}[t]
	\footnotesize
	\centering
	\resizebox{0.9\columnwidth}{!}{
		\begin{tabular}{p{5em}|cp{3em}cp{3.2em}cp{3em}cp{2.8em}}
			\toprule[1pt]
			\multicolumn{5}{c}{}
			\tabularnewline
			Activation & LT ($\mu s$) & AUCROC & AUCPR & ACC\tabularnewline
			\hline
			\hline
			Mish & 365 & $98.14_{2.04}$&$98.88_{1.13}$&$96.87_{2.46}$\tabularnewline
			Swish &  358 & $97.55_{1.90}$ & $98.39_{1.29}$&$96.39_{2.00}$ \tabularnewline 
			\bottomrule[1pt]
		\end{tabular}
	}
	\caption{Ablation on the effects of activation functions. Two popular activation functions are investigated for ScatterFormer on BECTS/Rolandic dataset. No significant difference is observed.}
	\label{tb:activations}
\end{table}
\begin{table}[t]
	\footnotesize
	\centering
	\resizebox{0.9\columnwidth}{!}{
		\centering
		\begin{tabular}{l|c c l l}
			\toprule[1pt]
			\multicolumn{5}{c}{\textbf{(a) BECTS/Rolandic Dataset}}
			\tabularnewline
			Model & Params (M) &   LT ($\mu s$)& AUCROC & ACC  \tabularnewline
			\hline
			\hline
			Scatter& 42& 365&$98.14_{2.04}$& $96.87_{2.46}$\tabularnewline
			Fourier& 39 & 239 & $96.97_{1.83}$ & $92.06_{2.96}$\tabularnewline 
			\bottomrule[1pt]
		\end{tabular}
	}
	\resizebox{0.9\columnwidth}{!}{
		\begin{tabular}{l|c c l l}
			\toprule[1pt]
			\multicolumn{5}{c}{\textbf{(b) Helsinki University Dataset}}
			\tabularnewline
			Model & Params (M)&  LT ($\mu s$)& AUCROC & ACC \tabularnewline
			\hline
			\hline
			Scatter & 42&  365&$96.38_{8.27}$&$90.05_{11.14}$\tabularnewline
			Fourier & 39 & 239 & $94.52_{8.92}$ & $89.04_{11.93}$\tabularnewline 
			\bottomrule[1pt]
	\end{tabular}}
	\caption{The Frequency-aware attention (FAA) module is replaced with Fast Fourier convolution to ablate its effects. The invariant scattering convolution outperforms the former by a median accuracy of 1.01\% with respect to neonatal seizure detection task (b) without significant increase in parameters. The unit of latency is $\mu s$.}
	\label{tb:ablation_attn}
\end{table}
\subsection{Ablation Studies}
\begin{figure*}[!t]
	\subfloat[ Raw EEG Waveform]
	{\includegraphics[width=0.39\linewidth]{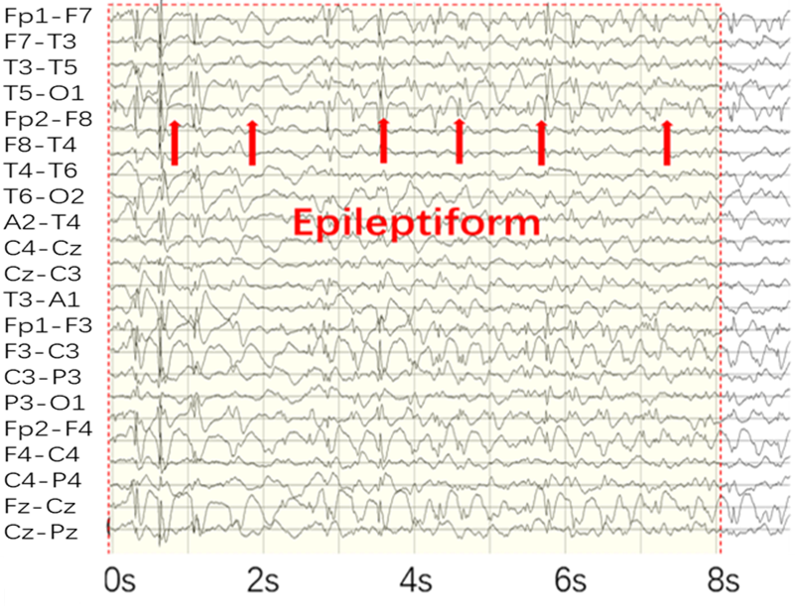}}
	\subfloat[Input]
	{
		\includegraphics[width=0.19\linewidth]{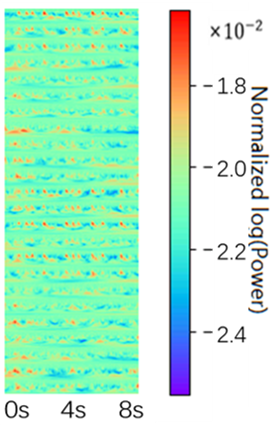}
	}
	\subfloat[Scatter]
	{\includegraphics[width=0.1\linewidth]{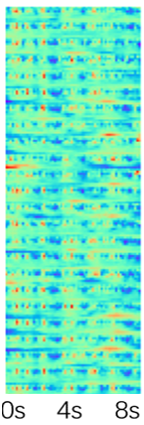}}
	\subfloat[Fourier]
	{\includegraphics[width=0.1\linewidth]{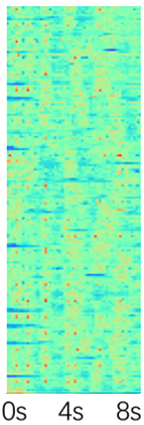}}
	\subfloat[Proto]
	{\includegraphics[width=0.198\linewidth]{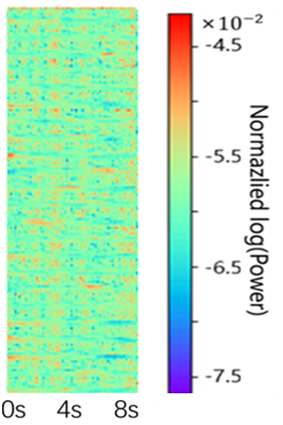}}
	\caption{Visualization of an EEG segment of epileptiform abnormalities and its corresponding visual representations. \textbf{(a)}: An illustrative EEG segment that contains continuous spike-and-wave during sleep (CSWS) characteristic of BECTS/Rolandic epilepsy. Epileptiform patterns are marked by red arrows for one channel with typical epileptic waves. \textbf{(b)}: Multispectral features. The color bar indicates normalized log wavelet power spectrum. \textbf{(c) (d) (e)}: Intermediate features learned by various Transformers. ScatterFormer (c) preserves more local spectral details that corresponds to spike-and-wave and sharp wave patterns appeared in the EEG than FourierFormer (d) and  ProtoFormer (e). The latter both suffer oversmoothing of high-frequency information to various extents. The color bar indicates normalized intermediate feature values that do not represent power spectrum of raw data.}
	\label{fig:comparison}
\end{figure*}
\subsubsection{Effects of Invariant Scattering Layer}
To validate the efficacy of the frequency-aware attention dependent on dual-branch token embedding mechanism that simultaneously utilizes convolution and invariant scattering transform, we investigate its role in learning. Detection accuracy is increased by 3.52\% compared to ProtoFormer (Table\;\ref{tb:results_1}). The substitution by fast Fourier convolution results in drop in performance (Table\;\ref{tb:ablation_attn}) despite slightly reducing the number of parameters.
\subsubsection{Effects of Activation}
ScatterFormer uses Mish activation. To investigate the effects of activation function, Mish is replaced with Swish, which leads to only negligible drops in several clinically important evaluation metrics with similar latency, as indicated in Table \;\ref{tb:activations}. Choice of nonlinear activation function does not seem to have significant influence on the expressivity and inference time of the network.
\subsection{Interpretability} 
In this section, we investigate the interpretability and robustness of ScatterFormer. In particular, we discuss the clinical translatability of the associated research results.

\subsubsection{Disentanglement of Attention}
To empirically corroborate the effectiveness of dual-branch design, we visualize the Fourier heat maps of high- and low-frequency branches of attention module, respectively, and quantitatively analyze the magnitude-frequency relationship. Observably, disentangled attention achieves slower decay of magnitude spectrum as frequency increases than the prototypical architecture and captures more high-frequency information signals (Figure 3). The response curve suggests concentration of energy primarily in low-frequency range, which is consistent with the results revealed in \cite{wang2022anti}. Noticeably, invariant scattering transform achieves more competitive performance than Fourier transform (Table\;\ref{tb:ablation_attn}). This performance gap could be attributed to energy preservation in scattering propagation, which corresponds to narrower frequency bands \cite{bruna2013invariant} and could potentially avoid oversmoothing while alleviating noises.
\subsubsection{Clinical Significance}

In Figure \;\ref{fig:comparison}, we inspect the feature maps learned by different variants to analyze which features were actually used for discriminative prediction. The results suggest that ScatterFormer learns more fine-grained spectral representations of eletrophysiological biomarkers such as spike-and-waves and sharp waves. Features generated by ProtoFormer are smoother.

In addition, fast Fourier convolution could lead to aliasing of distinguishing spectral patterns that correspond to epileptic spikes, and between channels, as shown in Figure \;\ref{fig:comparison}, which may explain the deteriorated inference-time performance. ScatterFormer enables more low-level features, especially regions associated with higher spectral power, to be captured. Noise and artifacts are suppressed at initial phase. It is suggested that our approach achieves better cross-subject accuracy because it works in a manner similar to expert electroencephalographers, who distinguish epileptiform abnormalities from other clinically irrelevant activities by detecting of fine-grained edge-like patterns.
\section{Conclusion}
In this work, we propose Scattering Transformer (ScatterFormer), an invariant scattering transform-based hierarchical Transformer that distinguishes subtle variations associated with high-frequency textural information for detecting epileptiform discharges. We theoretically prove that wavelet transform can increase features correlations to compensate the high-frequency features in transformer. Strong feature correlations of scattering transform or Fourier transform lead to reduced upper bound of Gaussian complexity, thus may improve model generalizability. Improvement of generalizability by transformation of features to scattering domain is further validated in experiments. 

Moreover, the scattering energy preservation feature allows more high-frequency information in different spatial, time, and frequency scales of epileptic EEG to be faithfully represented, which is reinforced by frequency-aware attention (FAA) that disentangles high- and low-frequency components. Therefore, the approach is able to differentiate subtle differences between spectral features converted from cEEG waveform records. We achieve optimal prediction AUCROC and accuracy in cross-subject detection epileptiform discharges in patients with BECTS/Rolandic epilepsy and neonates with heterogeneous etiologies for seizure, suggesting promise in accelerating clinical decision-making in various scenarios. Further analysis demonstrates the capability of ScatterFormer to extract discriminative patterns that are associated with epilepsy-specific EEG abnormalities, thereby offering clinical interpretability for supporting early and accurate identification of seizures. Our code is available at https://github.com/albertcheng19/scatterformer.
\section{Acknowledgments}
This work was supported from the Science and Technology Innovation 2030 - Brain Science and Brain-Inspired Intelligence Project (2021ZD0201301 and 2021ZD0200204), the National Natural Science Foundation of China (U20A20221), the Shanghai Municipal Science and Technology Major Project (2018SHZDZX01 and 2021SHZDZX0103) and ZJLab, Shanghai Municipal Science and Technology Committee of Shanghai outstanding academic leaders plan (21XD1400400).

\bibliography{aaai23}

\begin{thebibliography}{39}
\providecommand{\natexlab}[1]{#1}

\bibitem[{Achilles et~al.(2018)Achilles, Tombari, Belagiannis, Loesch,
  Noachtar, and Navab}]{achilles2018convolutional}
Achilles, F.; Tombari, F.; Belagiannis, V.; Loesch, A.~M.; Noachtar, S.; and
  Navab, N. 2018.
\newblock Convolutional neural networks for real-time epileptic seizure
  detection.
\newblock \emph{Computer Methods in Biomechanics and Biomedical Engineering:
  Imaging \& Visualization}, 6(3): 264--269.

\bibitem[{Ali et~al.(2021)Ali, Touvron, Caron, Bojanowski, Douze, Joulin,
  Laptev, Neverova, Synnaeve, Verbeek et~al.}]{ali2021xcit}
Ali, A.; Touvron, H.; Caron, M.; Bojanowski, P.; Douze, M.; Joulin, A.; Laptev,
  I.; Neverova, N.; Synnaeve, G.; Verbeek, J.; et~al. 2021.
\newblock Xcit: Cross-covariance image transformers.
\newblock \emph{Advances in neural information processing systems}, 34:
  20014--20027.

\bibitem[{Asif et~al.(2020)Asif, Roy, Tang, and Harrer}]{asif2020seizurenet}
Asif, U.; Roy, S.; Tang, J.; and Harrer, S. 2020.
\newblock SeizureNet: Multi-spectral deep feature learning for seizure type
  classification.
\newblock In \emph{Machine Learning in Clinical Neuroimaging and Radiogenomics
  in Neuro-oncology}, 77--87. Springer.

\bibitem[{Bagchi and Bathula(2022)}]{bagchi2022eeg}
Bagchi, S.; and Bathula, D.~R. 2022.
\newblock EEG-ConvTransformer for single-trial EEG-based visual stimulus
  classification.
\newblock \emph{Pattern Recognition}, 129: 108757.

\bibitem[{Bai et~al.(2022)Bai, Yuan, Xia, Yan, Li, and Liu}]{bai2022improving}
Bai, J.; Yuan, L.; Xia, S.-T.; Yan, S.; Li, Z.; and Liu, W. 2022.
\newblock Improving Vision Transformers by Revisiting High-frequency
  Components.
\newblock \emph{arXiv preprint arXiv:2204.00993}.

\bibitem[{Bruna and Mallat(2013)}]{bruna2013invariant}
Bruna, J.; and Mallat, S. 2013.
\newblock Invariant scattering convolution networks.
\newblock \emph{IEEE transactions on pattern analysis and machine
  intelligence}, 35(8): 1872--1886.

\bibitem[{Cotter and Kingsbury(2019)}]{cotter2019learnable}
Cotter, F.; and Kingsbury, N. 2019.
\newblock A learnable scatternet: Locally invariant convolutional layers.
\newblock In \emph{2019 IEEE International Conference on Image Processing
  (ICIP)}, 350--354. IEEE.

\bibitem[{Dong et~al.(2021)Dong, Bao, Chen, Zhang, Yu, Yuan, Chen, and
  Guo}]{dong2021cswin}
Dong, X.; Bao, J.; Chen, D.; Zhang, W.; Yu, N.; Yuan, L.; Chen, D.; and Guo, B.
  2021.
\newblock Cswin transformer: A general vision transformer backbone with
  cross-shaped windows.
\newblock \emph{arXiv preprint arXiv:2107.00652}.

\bibitem[{Frassineti et~al.(2020)Frassineti, Ermini, Fabbri, and
  Manfredi}]{9140713}
Frassineti, L.; Ermini, D.; Fabbri, R.; and Manfredi, C. 2020.
\newblock Neonatal Seizures Detection using Stationary Wavelet Transform and
  Deep Neural Networks: Preliminary Results.
\newblock In \emph{2020 IEEE 20th Mediterranean Electrotechnical Conference (
  MELECON)}, 344--349.

\bibitem[{Isaev et~al.(2020)Isaev, Tchapyjnikov, Cotten, Tanaka, Martinez,
  Bertran, Sapiro, and Carlson}]{isaev2020attention}
Isaev, D.~Y.; Tchapyjnikov, D.; Cotten, C.~M.; Tanaka, D.; Martinez, N.;
  Bertran, M.; Sapiro, G.; and Carlson, D. 2020.
\newblock Attention-based network for weak labels in neonatal seizure
  detection.
\newblock \emph{Proceedings of machine learning research}, 126: 479.

\bibitem[{Jeong et~al.(2021)Jeong, Jeon, Ko, and Suk}]{9385307}
Jeong, S.; Jeon, E.; Ko, W.; and Suk, H.-I. 2021.
\newblock Fine-grained Temporal Attention Network for EEG-based Seizure
  Detection.
\newblock In \emph{2021 9th International Winter Conference on Brain-Computer
  Interface (BCI)}, 1--4.

\bibitem[{Li et~al.(2020)Li, Gao, Zhang, Huang, Wu, and
  Xu}]{li2020distinguishing}
Li, Q.; Gao, J.; Zhang, Z.; Huang, Q.; Wu, Y.; and Xu, B. 2020.
\newblock Distinguishing epileptiform discharges from normal
  electroencephalograms using adaptive fractal and network analysis: a clinical
  perspective.
\newblock \emph{Frontiers in Physiology}, 828.

\bibitem[{Li et~al.(2017)Li, Li, Fern, and Raich}]{li2017filter}
Li, X.; Li, F.; Fern, X.; and Raich, R. 2017.
\newblock Filter Shaping for Convolutional Neural Network.
\newblock In \emph{International Conference on Learning Representations}.

\bibitem[{Liu et~al.(2021)Liu, Lin, Cao, Hu, Wei, Zhang, Lin, and
  Guo}]{liu2021swin}
Liu, Z.; Lin, Y.; Cao, Y.; Hu, H.; Wei, Y.; Zhang, Z.; Lin, S.; and Guo, B.
  2021.
\newblock Swin transformer: Hierarchical vision transformer using shifted
  windows.
\newblock In \emph{Proceedings of the IEEE/CVF International Conference on
  Computer Vision}, 10012--10022.

\bibitem[{Mallat(2012)}]{mallat2012group}
Mallat, S. 2012.
\newblock Group invariant scattering.
\newblock \emph{Communications on Pure and Applied Mathematics}, 65(10):
  1331--1398.

\bibitem[{Naseer et~al.(2021)Naseer, Ranasinghe, Khan, Hayat, Shahbaz~Khan, and
  Yang}]{NEURIPS2021_c404a5ad}
Naseer, M.~M.; Ranasinghe, K.; Khan, S.~H.; Hayat, M.; Shahbaz~Khan, F.; and
  Yang, M.-H. 2021.
\newblock Intriguing Properties of Vision Transformers.
\newblock In Ranzato, M.; Beygelzimer, A.; Dauphin, Y.; Liang, P.; and Vaughan,
  J.~W., eds., \emph{Advances in Neural Information Processing Systems},
  volume~34, 23296--23308. Curran Associates, Inc.

\bibitem[{Oyallon et~al.(2018)Oyallon, Belilovsky, Zagoruyko, and
  Valko}]{oyallon2018compressing}
Oyallon, E.; Belilovsky, E.; Zagoruyko, S.; and Valko, M. 2018.
\newblock Compressing the input for cnns with the first-order scattering
  transform.
\newblock In \emph{Proceedings of the European Conference on Computer Vision
  (ECCV)}, 301--316.

\bibitem[{Pan, Cai, and Zhuang(2022)}]{pan2022fast}
Pan, Z.; Cai, J.; and Zhuang, B. 2022.
\newblock Fast Vision Transformers with HiLo Attention.
\newblock \emph{arXiv preprint arXiv:2205.13213}.

\bibitem[{Radosavovic et~al.(2020)Radosavovic, Kosaraju, Girshick, He, and
  Doll{\'a}r}]{radosavovic2020designing}
Radosavovic, I.; Kosaraju, R.~P.; Girshick, R.; He, K.; and Doll{\'a}r, P.
  2020.
\newblock Designing network design spaces.
\newblock In \emph{Proceedings of the IEEE/CVF Conference on Computer Vision
  and Pattern Recognition}, 10428--10436.

\bibitem[{Rasheed et~al.(2020)Rasheed, Qayyum, Qadir, Sivathamboo, Kwan,
  Kuhlmann, O’Brien, and Razi}]{rasheed2020machine}
Rasheed, K.; Qayyum, A.; Qadir, J.; Sivathamboo, S.; Kwan, P.; Kuhlmann, L.;
  O’Brien, T.; and Razi, A. 2020.
\newblock Machine learning for predicting epileptic seizures using EEG signals:
  A review.
\newblock \emph{IEEE Reviews in Biomedical Engineering}, 14: 139--155.

\bibitem[{Saminu et~al.(2021)Saminu, Xu, Shuai, Abd El~Kader, Jabire, Ahmed,
  Karaye, and Ahmad}]{saminu2021recent}
Saminu, S.; Xu, G.; Shuai, Z.; Abd El~Kader, I.; Jabire, A.~H.; Ahmed, Y.~K.;
  Karaye, I.~A.; and Ahmad, I.~S. 2021.
\newblock A Recent Investigation on Detection and Classification of Epileptic
  Seizure Techniques Using EEG Signal.
\newblock \emph{Brain Sciences}, 11(5): 668.

\bibitem[{Sazgar and Young(2019)}]{sazgar2019overview}
Sazgar, M.; and Young, M.~G. 2019.
\newblock Overview of EEG, electrode placement, and montages.
\newblock In \emph{Absolute epilepsy and EEG rotation review}, 117--125.
  Springer.

\bibitem[{Shi et~al.(2020)Shi, Zhang, Miao, Sun, Sun, Chen, Hu, Xiang, and
  Wang}]{shi2020differences}
Shi, Q.; Zhang, T.; Miao, A.; Sun, J.; Sun, Y.; Chen, Q.; Hu, Z.; Xiang, J.;
  and Wang, X. 2020.
\newblock Differences between interictal and ictal generalized spike-wave
  discharges in childhood absence epilepsy: a MEG study.
\newblock \emph{Frontiers in Neurology}, 1359.

\bibitem[{Shorvon and Schmidt(2016)}]{shorvon2016right}
Shorvon, S.; and Schmidt, D. 2016.
\newblock The right and the wrong with epilepsy and her science.
\newblock \emph{Epilepsia Open}, 1(3-4): 76--85.

\bibitem[{Si et~al.(2022)Si, Yu, Zhou, Zhou, Wang, and Yan}]{si2022inception}
Si, C.; Yu, W.; Zhou, P.; Zhou, Y.; Wang, X.; and Yan, S. 2022.
\newblock Inception Transformer.
\newblock \emph{arXiv preprint arXiv:2205.12956}.

\bibitem[{Siddhad et~al.(2022)Siddhad, Gupta, Dogra, and
  Roy}]{siddhad2022efficacy}
Siddhad, G.; Gupta, A.; Dogra, D.~P.; and Roy, P.~P. 2022.
\newblock Efficacy of Transformer Networks for Classification of Raw EEG Data.
\newblock \emph{arXiv preprint arXiv:2202.05170}.

\bibitem[{Singh and Kingsbury(2017)}]{singh2017dual}
Singh, A.; and Kingsbury, N. 2017.
\newblock Dual-tree wavelet scattering network with parametric log
  transformation for object classification.
\newblock In \emph{2017 IEEE International Conference on Acoustics, Speech and
  Signal Processing (ICASSP)}, 2622--2626. IEEE.

\bibitem[{Singh and Kingsbury(2018)}]{singh2018generative}
Singh, A.; and Kingsbury, N. 2018.
\newblock Generative scatternet hybrid deep learning (g-shdl) network with
  structural priors for semantic image segmentation.
\newblock In \emph{2018 IEEE International Conference on Acoustics, Speech and
  Signal Processing (ICASSP)}, 2991--2995. IEEE.

\bibitem[{Stevenson et~al.(2019)Stevenson, Tapani, Lauronen, and
  Vanhatalo}]{stevenson2019dataset}
Stevenson, N.~J.; Tapani, K.; Lauronen, L.; and Vanhatalo, S. 2019.
\newblock A dataset of neonatal EEG recordings with seizure annotations.
\newblock \emph{Scientific data}, 6(1): 1--8.

\bibitem[{Sun, Xie, and Zhou(2021)}]{sun2021eeg}
Sun, J.; Xie, J.; and Zhou, H. 2021.
\newblock EEG classification with transformer-based models.
\newblock In \emph{2021 IEEE 3rd Global Conference on Life Sciences and
  Technologies (LifeTech)}, 92--93. IEEE.

\bibitem[{Tao et~al.(2021)Tao, Sun, Muhamed, Genc, Jackson, Arsanjani,
  Yaddanapudi, Li, and Kumar}]{tao2021gated}
Tao, Y.; Sun, T.; Muhamed, A.; Genc, S.; Jackson, D.; Arsanjani, A.;
  Yaddanapudi, S.; Li, L.; and Kumar, P. 2021.
\newblock Gated transformer for decoding human brain eeg signals.
\newblock In \emph{2021 43rd Annual International Conference of the IEEE
  Engineering in Medicine \& Biology Society (EMBC)}, 125--130. IEEE.

\bibitem[{Tatum et~al.(2018)Tatum, Rubboli, Kaplan, Mirsatari, Radhakrishnan,
  Gloss, Caboclo, Drislane, Koutroumanidis, Schomer et~al.}]{tatum2018clinical}
Tatum, W.; Rubboli, G.; Kaplan, P.; Mirsatari, S.; Radhakrishnan, K.; Gloss,
  D.; Caboclo, L.; Drislane, F.; Koutroumanidis, M.; Schomer, D.; et~al. 2018.
\newblock Clinical utility of EEG in diagnosing and monitoring epilepsy in
  adults.
\newblock \emph{Clinical Neurophysiology}, 129(5): 1056--1082.

\bibitem[{Tatum(2012)}]{tatum2012eeg}
Tatum, W.~O. 2012.
\newblock EEG interpretation: common problems.
\newblock \emph{Clinical Practice}, 9(5): 527.

\bibitem[{Tatum and Shellhaas(2020)}]{Tatum862}
Tatum, W.~O.; and Shellhaas, R.~A. 2020.
\newblock Epileptiform discharges.
\newblock \emph{Neurology}, 94(20): 862--863.

\bibitem[{Wang et~al.(2022)Wang, Zheng, Chen, and Wang}]{wang2022anti}
Wang, P.; Zheng, W.; Chen, T.; and Wang, Z. 2022.
\newblock Anti-oversmoothing in deep vision transformers via the fourier domain
  analysis: From theory to practice.
\newblock \emph{arXiv preprint arXiv:2203.05962}.

\bibitem[{Wu et~al.(2021)Wu, Xiao, Codella, Liu, Dai, Yuan, and
  Zhang}]{wu2021cvt}
Wu, H.; Xiao, B.; Codella, N.; Liu, M.; Dai, X.; Yuan, L.; and Zhang, L. 2021.
\newblock Cvt: Introducing convolutions to vision transformers.
\newblock In \emph{Proceedings of the IEEE/CVF International Conference on
  Computer Vision}, 22--31.

\bibitem[{Xiang et~al.(2015)Xiang, Li, Li, Cao, Wang, Han, and
  Chen}]{xiang2015detection}
Xiang, J.; Li, C.; Li, H.; Cao, R.; Wang, B.; Han, X.; and Chen, J. 2015.
\newblock The detection of epileptic seizure signals based on fuzzy entropy.
\newblock \emph{Journal of neuroscience methods}, 243: 18--25.

\bibitem[{Yum and Shvarts(2019)}]{yum_shvarts_2019}
Yum, A.; and Shvarts, V. 2019.
\newblock \emph{Ictal and Interictal Epileptiform Electroencephalogram
  Patterns}, 239–250.
\newblock Cambridge University Press.

\bibitem[{Zhang et~al.(2017)Zhang, Cisse, Dauphin, and
  Lopez-Paz}]{zhang2017mixup}
Zhang, H.; Cisse, M.; Dauphin, Y.~N.; and Lopez-Paz, D. 2017.
\newblock mixup: Beyond empirical risk minimization.
\newblock \emph{arXiv preprint arXiv:1710.09412}.

\end{thebibliography}

\end{document}